\documentclass[rnote,traditabstract]{aa}  

\usepackage{graphicx}
\usepackage{txfonts}
\begin{document}
   \title{Distance and reddening of the Local Group dwarf irregular galaxy NGC 6822\thanks{Based on observations collected with the ACS onboard the NASA/ESA HST}}

\authorrunning{Fusco et al.}
\titlerunning{The LG dIrr galaxy NGC~6822}

   \author{F. Fusco     \inst{1}, R. Buonanno  \inst{1,2}, G. Bono \inst{1}, S. Cassisi \inst{2}, M. Monelli \inst{3,4}, \and A. Pietrinferni \inst{2}}

   \institute{Universit\`a di Roma Tor Vergata, Via della Ricerca Scientifica 1, 00133 Rome, Italy\\
              \email{federica.fusco@roma2.infn.it}
         \and
             INAF- Osservatorio Astronomico di Teramo, Via Mentore Maggini, 64100 Teramo, Italy  
         \and
            Instituto Astrofisico de Canarias, Calle Via Lactea, E38200 La Laguna, Tenerife, Spain
          \and
           Departamento de Astrof\'isica, Universidad de La Laguna, Tenerife, Spain
             }

   \date{Received  ; accepted }

 \abstract{On the basis of a new photometric analysis of the Local Group dwarf irregular galaxy NCG 6822 based on observations obtained with the Advanced Camera for Surveys onboard the Hubble Space Telescope, we have obtained a new estimate of the extinction of two fields located in the southeast region of the galaxy. Because of significant differences in the distance estimates to NGC~6822 available in literature, we decided to provide an independent determination of the distance to this galaxy based on an updated and self-consistent theoretical calibration of the tip of the red giant branch brightness. As a result we newly determined the distance to NGC~6822 to be equal to ${\rm(m-M)}_0=23.54\pm 0.05$, and compared our measurement with the most recent determinations of this distance.}
\keywords{galaxies: individual (NGC 6822)\textemdash Local Group\textemdash stars: distances} 
\maketitle 

%

\section{Introduction}
NGC 6822 is a barred dwarf irregular galaxy (dIrr) of type Ir IV-V (Gallart et al. 1996, van den Bergh \cite{vandenberg}) belonging to the Local Group (LG). This galaxy, discovered by E. E. Barnard (\cite{barnard}), is one of the nearest dIrr to the Milky Way (MW), and according to Cannon et al. (\cite{Cannon}) it is rich in gas and still actively forming stars. In addition, NGC 6822 is dark-matter-dominated (de Blok \& Walter \cite{deblok}). This LG dwarf galaxy is similar in size, structure, and metallicity to the Small Magellanic Cloud (SMC) and therefore could play a fundamental role in understanding the star formation history in complex stellar systems. 

Several estimates of the metal abundance are found in the literature. Skillman et al. (\cite{skillman}) estimated ${\rm [Fe/H]}=-1.2\pm 0.3$ through
optical spectrophotometry, Gallart et al. (\cite{gallart}) found ${\rm
[Fe/H]}=-1.5\pm 0.3$ using optical photometry, while through the CaII triplet absorption lines in red giant branch (RGB) stars
Tolstoy et al. (\cite{tolstoy}) give ${\rm [Fe/H]}=-1.0\pm 0.3$, Venn et al. (\cite{venn}) estimated ${\rm
[Fe/H]}=-0.49\pm 0.22$ using optical spectroscopy of A-type stars, and finally, using the C/M Asymptotic Giant Branch star star ratio Kacharov et al. (\cite{kacharov}) and Sibbons et al. (\cite{sibbons}) found ${\rm [Fe/H]}=-1.3\pm 0.2$ and ${\rm [Fe/H]}=-1.29\pm 0.07$, respectively. 

NGC 6822 is located at a low galactic latitude ($l=25.4^{\circ}, b=-18.4^{\circ}$, Mateo \cite{mateo}) and is therefore affected by a moderate foreground extinction. Indeed, according to Schelgel's maps (\cite{schlegel}), updated in Schlafly \& Finkbeiner (\cite{schlafly}), the MW reddening along the line of sight towards NGC 6822 is about ${\rm E(B-V)}=0.21$. The reddening estimates show a clear-cut difference between the innermost and outermost regions. Massey et al. (\cite{massey}) gave ${\rm E(B-V)}=0.26$ and ${\rm E(B-V)}=0.45$ in the centre and in the external fields of the galaxy; Gallart et al. (\cite{gallart}) obtained ${\rm E(B-V)}=0.24\pm 0.03$ in the outermost regions of the galaxy, and Gieren et al. (\cite{gieren}) estimated ${\rm E(B-V)}=0.356\pm 0.013$ in a region near the centre. 

The distance modulus of this dwarf irregular has been estimated by several authors. Lee et al. (\cite{lee})
and Salaris \& Cassisi (\cite{sc:98}) gave
${\rm(m-M)}_0=23.46\pm 0.08$ and $23.71\pm0.14$ using the RGB tip brightness as a distance indicator; Gallart et al. (\cite{gallart}) estimate ${\rm(m-M)}_0=23.49\pm 0.08$ using UBVRI
photometry of eight Cepheids; Cioni \& Habing (\cite{cioni}) using near-infrared (NIR) 
photometry of AGB stars found ${\rm(m-M)}_0=23.34\pm 0.12$; Gieren et al. (\cite{gieren}) found ${\rm(m-M)}_{0} = 23.312 \pm 0.021$ through NIR
photometry of Cepheids in the innermost region of NGC 6822.
More recently, significant differences among the distance estimates to NGC~6822, based on the use of the classical Cepheid period--luminosity relation, have been extensively investigated by Feast et al. (2012). These authors adopted a NIR JHK photometric dataset of classical Cepheids in the central region of NGC~6822 and obtained a (best-value) distance of ${\rm(m-M)}_0=23.40\pm 0.05$ based on a Large Magellanic Cloud (LMC) distance modulus of 18.50 mag, and a mean visual extinction ${\rm A_V=0.667}$. 
A new independent estimate of the distance to NGC~6822, based on the NIR period--luminosity relation of Miras variables and assuming an LMC distance of 18.50, has also been obtained by Whitelock et al.~(2012): $23.56\pm0.03$, where the error does not account for the uncertainty in the LMC distance.

Because it is crucial to obtain accurate distance determinations for LG dwarfs and test the self-consistency of independent standard candles (Tammann, Sandage \& Reindl~2008), we decided to 
obtain a new determination of the distance to NGC~6822 based on the the tip of the Red Giant Branch (TRGB) method by adopting a new theoretical calibration of this distance indicator. In passing we note that while the distance measurements based on classical Cepheids are related to the young and intermediate-age galactic stellar component, 
the distance estimates provided by the TRGB method are tightly related to the old stellar population in the galaxy.


\section{Observations and data reduction}
Our photometric catalog is based on archival {\textit{Hubble Space Telescope}} ({\textit{HST}}) data sets collected with the Advanced Camera for Surveys (ACS) (HST proposal GO-12180 P.I. J. M. Cannon). We present here the analysis of three fields, all located in the southeast region of the galaxy\footnote{The centre of NGC 6822 is $(\alpha,\delta)=(-19h44m56.6s, -14d47m21s)$}. The three fields were labelled C, E1, and E2, from the inside to the outside. Field, filter, exposure, and coordinates of our data set are listed in Table \ref{table:fields}.

%
%
\begin{table}
\caption{List of images in the three fields, with filter, exposure time (s), $\alpha$ (h), and $\delta$ (deg).}             
\label{table:fields}      
\centering          
\begin{tabular}{c c c c c}     
\hline\hline       
                      
Field & Filter & Exp. Time (s) & $\alpha$ (h) & $\delta$ (deg)\\
\hline                    
C  & F475W &  559.5 &  19.753  &  -14.829\\
C  & F475W &  559.5 &  19.753  &  -14.829\\
C  & F814W &  559.0 &  19.753  &  -14.829\\
C  & F814W &  559.0 &  19.753  &  -14.829\\
E1 & F475W &  559.5 &  19.758  &  -14.887\\
E1 & F475W &  559.5 &  19.758  &  -14.887\\
E1 & F814W &  559.0 &  19.758  &  -14.887\\
E1 & F814W &  559.0 &  19.758  &  -14.887\\
E2 & F475W &  423.0 &  19.764  &  -14.943\\
E2 & F475W &  463.0 &  19.764  &  -14.943\\
E2 & F814W &  883.0 &  19.764  &  -14.943\\
E2 & F814W &  463.0 &  19.764  &  -14.943\\

\hline                  
\end{tabular}
\end{table}

As a first step, we corrected the images for the known distortion effects using the pixel area map (PAM). Photometry on individual images was
performed using the photometric package DAOPHOT (Stetson \cite{daophot}), and the simultaneous photometry on
the four images was carried out using ALLFRAME (Stetson \cite{allframe}). We assembled three catalogues
\textemdash one for each field\textemdash\ containing $\sim$140,000, $\sim$42,000, and
$\sim$12,000 stars, from the innermost to the outermost field. The data calibration was carried out using the VEGAMAG system (Sirianni et al. \cite{sirianni}) once adopting the most updated zero points\footnote{Available at http://www.stsci.edu/hst/acs/analysis/zeropoints.}.

\begin{figure}
   \centering
   \includegraphics[scale=0.5]{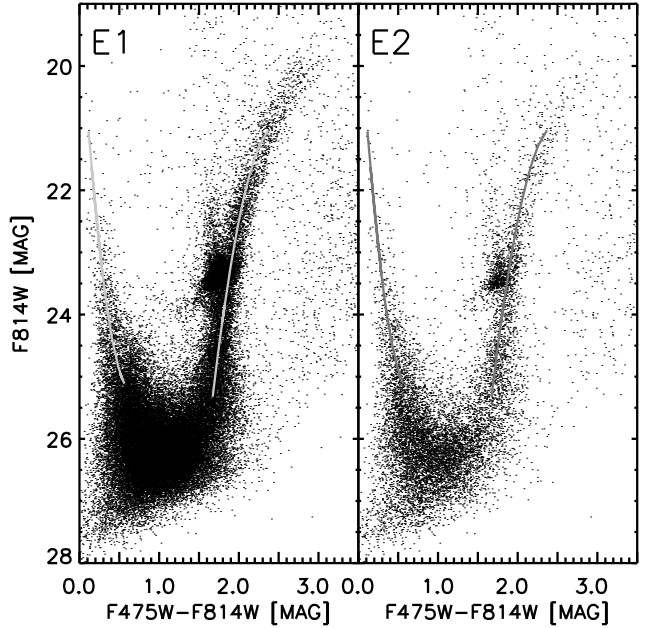}
   \caption{{\textit {Left}}: CMD of the E1 region with the ridge-lines superimposed. {\textit {Right}}: CMD of field E2 with the ridge-lines of E1 superimposed. The values of the ridge-lines are listed in Tab.~\ref{table:ridgelines}.}
    \label{cmd_lin}%
    \end{figure}

We thus obtained three F814W, F475W$-$F814W colour magnitude diagrams (CMDs), one for each field. Before analysing the three CMDs in more detail, we focused on the CMDs of the external fields, E1 and E2. Considering that each of the two fields contained a relatively small number of stars, we exploited the possibility to increase the statistics by merging the two catalogues. First, we computed the ridge-lines of the main sequence (MS) and the RGB for field E1 and superimposed them onto the CMD of field E2. Fig.~\ref{cmd_lin} shows the resulting excellent overlap. Therefore, to study a sample of higher statistical significance, we merged the CMDs of fields E1 and E2 and hereafter only a single CMD \textemdash labelled E\textemdash\ is considered as representative of the external population of NGC 6822.

\begin{figure}
   \centering
   \includegraphics[scale=0.5]{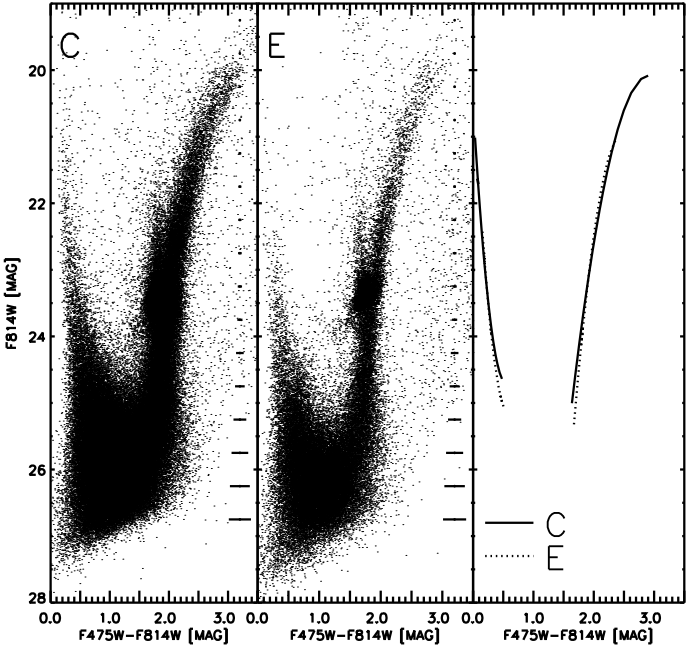}
   \caption{{\textit{Left}}: CMD of the innermost field, C. {\textit {Middle}}: The CMD of the outermost fields, E. The error bars are shown. {\textit {Right}}: The ridge-lines of the MS and of the RGB of the fields C (solid line) and E (dot line). The ridge-lines of field C blue shifted by ${\rm E(F475W-F814W)}=0.12$ and shifted to faint magnitudes by $\Delta m=0.12$ (see text).}
              \label{cmd_final}%
    \end{figure}

The two CMDs of field C and E are shown in the left and middle panel of Fig.~\ref{cmd_final} including the error bars at different magnitudes.

The last step was to find the relative shift in colour (and in magnitude) of the two CMDs of fields C and E. After tracing the ridge-lines of the two CMDs, we found an excellent superposition for 
\begin{equation}
    \Delta {\rm E(F475W-F814W)}= 0.12\pm 0.02 
\label{eq:diff_red}
\end{equation}
and $\Delta m=0.12\pm 0.02$ \textemdash hereafter {\textit m} indicates the apparent magnitude in F814W band\textemdash, as shown in Fig.~\ref{cmd_final}. In other words, an excellent agreement is found assuming that field C is 0.12 redder and fainter than the CMD of field E. The associated error simply reflects the uncertainty in tracing the ridge-lines because of the intrinsic dispersion of the sequences in the CMD.

   \begin{figure}
   \centering
   \includegraphics[scale=0.5]{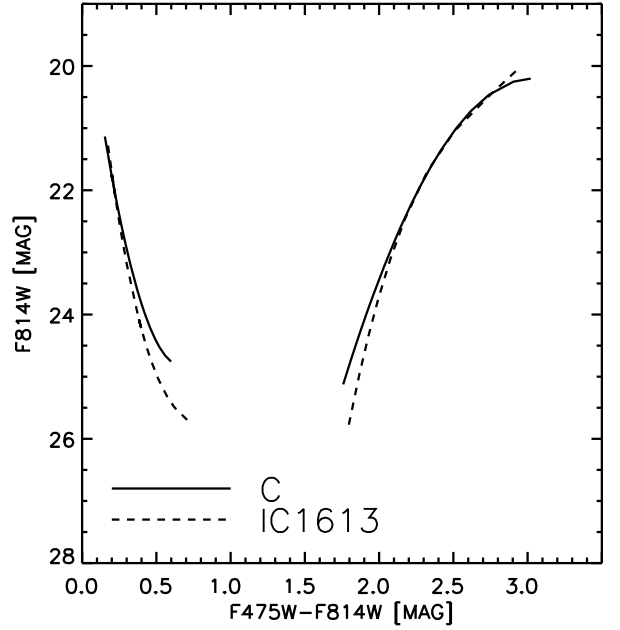}
\caption{Ridge-lines of the MS and RGB of the fields C (solid line) and IC 1613 (dashed line). The colour shift applied to the ridge-lines of the two galaxies to overlap them  corresponds to  ${\rm \Delta{(F475W-F814W)}=0.59\pm0.02}$.}
 \label{cmd_diff_red}
    \end{figure}

\section{Distance and reddening determination}

To provide a new estimate of the reddening of NGC 6822, we began by adopting a classical approach. In particular we used the CMD of the dIrr galaxy IC 1613, whose CMD in the HST bands (Bernard et al. \cite{bernard}) presents a global morphology very similar to that of NGC 6822. The similarity of the CMDs suggests that the two galaxies experienced an analogue star formation history. In addition, the reddening of IC 1613 is as low as ${\rm E(B-V)} \sim 0.026$ (Bernard et al. \cite{bernard}). After determining the ridge-lines of the MS and the RGB of IC 1613 (ridge-lines for the MS and the RGB of fields C and E of NGC 6822 and of IC 1613 are reported in Table~\ref{table:ridgelines}), we applied an appropriate shift both in magnitude and colour to obtain the best overlapping of the ridge-lines. This is shown in Fig.~\ref{cmd_diff_red}. As a result of this procedure, we obtained ${\rm \Delta{(F475W-F814W)}=0.59\pm0.02}$ and $\Delta m=1.00\pm0.04$ as the relative colour and magnitude of the NGC 6822 field C with respect to IC 1613.

On the basis of studying the rich population of variables in IC 1613 Bernard et al. (\cite{bernard}) found for the distance modulus of this galaxy ${\rm(m-M)_0=24.40}$ with an uncertainty (our estimate) of $\sim$0.05. Therefore we estimate the distance modulus of NGC 6822 as ${\rm(m-M)_0=23.40\pm0.06}$.

\begin{table}
\caption{Magnitudes in the F814W band associated to each colour F475W$-$F814W for the ridge-lines of fields C, E, and for IC 1613.}             
\label{table:ridgelines}      
\centering          
\begin{tabular}{c c c c c c c}     
\hline\hline       
                      
Colour        &   $m_C^{MS}$   & $ m_E^{MS}$   &  $m_{IC 1613}^{MS}$  &  $m_C^{RGB}$   & $ m_E^{RGB}$   &  $m_{IC 1613}^{RGB}$\\
\hline                    
      2.5     &   --           &    --         &            --        &     21.17      &     20.07      &    20.95 \\
      2.4     &   --           &    --         &            --        &     21.46      &     20.38      &    20.99 \\
      2.3     &   --           &    --         &            --        &     21.81      &     20.64      &    21.17 \\
      2.2     &   --           &    --         &            --        &     22.24      &     20.88      &    21.48 \\
      2.1     &   --           &    --         &            --        &     22.75      &     21.11      &    21.92 \\
      2.0     &   --           &    --         &            --        &     23.35      &     21.36      &    22.49 \\
      1.9     &   --           &    --         &            --        &     24.07      &     21.64      &    23.20 \\
      1.8     &   --           &    --         &            --        &     24.91      &     21.98      &    24.05 \\
      0.5     &     24.43      &     26.54     &          24.90       &        --      &       --       &     --   \\
      0.4     &     23.83      &     26.47     &          24.33       &        --      &       --       &     --   \\
      0.3     &     22.94      &     26.41     &          23.46       &        --      &       --       &     --   \\
      0.2     &     21.78      &     26.28     &          22.29       &        --      &       --       &     --   \\
      0.1     &     20.33      &     26.02     &          20.83       &        --      &       --       &     --   \\
      0.0     &     18.61      &     25.56     &          19.07       &        --      &       --       &     --   \\
     -0.1     &     16.61      &     24.84     &          17.02       &        --      &       --       &     --   \\


\hline                  
\end{tabular}
\end{table}

An independent estimate of the distance to NGC 6822 can be obtained using the luminosity of the RGB tip (TRGB) (Lee et al. \cite{lee}). The basic idea of this technique is that the TRGB marks the He ignition in electron degenerate cores of low-mass stars, i.e. stars with masses below ${\rm \sim2M_{\odot}}$, corresponding to population ages older than 1.0-1.5~Gyr. 

\begin{figure}
\centering
\includegraphics[scale=0.5]{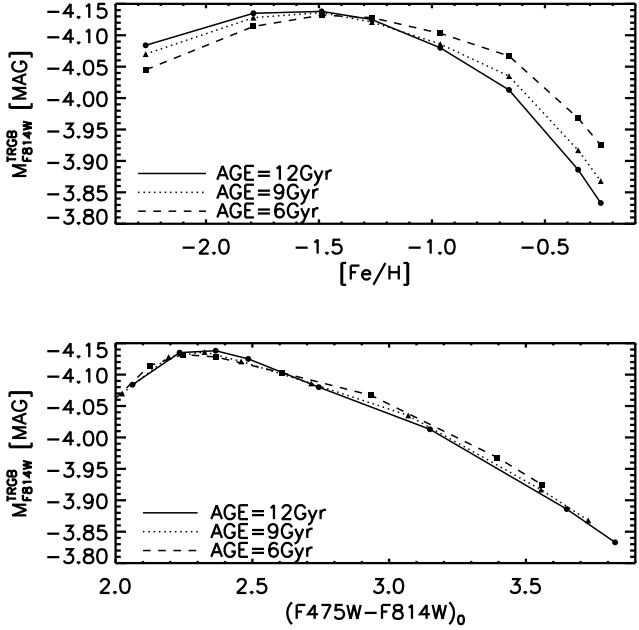}
 \caption{{\sl Top panel}: the luminosity of the TRGB in the F814W-band as a function of the metallicity for different ages (12 Gyr: solid line; 9 Gyr: dotted line; 6 Gyr: dashed line). {\sl Bottom panel}: the luminosity of the TRGB as  function of the colour on the TRGB for different ages (12 Gyr: solid line; 9 Gyr: dotted line; 6 Gyr: dashed line). Different metallicities are indicated by the small dots.}
\label{trgb_models}
\end{figure}

The TRGB bolometric magnitude is weakly dependent on the initial stellar mass for ages older than 4~Gyr because, for a given initial chemical composition, the TRGB level is determined by the He-core mass at the He flash, which is fairly constant in this age range. The core mass at the He-burning ignition decreases for increasing metallicity, while the TRGB bolometric luminosity increases. This is so because the increased efficiency of the H-burning shell largely compensates for the reduced core mass. However, according to the models of stellar atmosphere, the bolometric correction in the I-Cousin band nearly compensates for the intrinsic increase in the bolometric luminosity of the TRGB. The net effect is that ${\rm M_I^{TRGB}}$ should be largely constant with ${\rm M_I^{TRGB}\approx-4}$ and weakly dependent on ages and metallicities for ages older than 4-5~Gyr and metallicities lower than [Fe/H]$\leq$-0.7  (Lee et al.~\cite{lee}, Salaris \& Cassisi~\cite{sc:05}). 
These model predictions are shown in the upper panel of Fig.~\ref{trgb_models}. For this reason the calibration of the TRGB brightness in the I-band (or in similar bands such as F814W) as a function of the metallicity can be used as a standard candle for resolved old stellar populations.

However, even if the TRGB is an extremely powerful distance indicator for homogeneous stellar populations,
when applied to multi-age stellar systems -- such as those hosted in LG dwarfs-- it should be used with caution.
The problem is that even if weak, the dependence on the age of the TRGB is not completely negligible.
Consequently, a poor knowledge of the star formation history could affect the estimate of the distance, as
discussed by Tammann \& Reindl (\cite{tammann}). The presence of a well-populated RGB in the CMD of
a stellar system with a complex formation history does not guarantee that the RGB is populated only by old
stars. To overcome this difficulty, several studies, both observational (Rizzi et al. \cite{rizzi}) and theoretical (Bellazzini \cite{bellazzini}, Cassisi \& Salaris \cite{cs:2012}) have been carried out and concluded that the calibration of the TRGB absolute magnitude as a function of its intrinsic colour has the advantage of minimising the effect of the star formation history of the star population. 

On the basis of these considerations, we decided to estimate the distance to NGC 6822 by adopting the calibration of the TRGB brightness as a function of both the metallicity and intrinsic colour.

Concerning the theoretical calibration of the TRGB, we relied on the BaSTI stellar models (Pietrinferni et al. \cite{pietrinferni04, pietrinferni06}), updated with conductive opacity evaluations by Cassisi et al. (\cite{cassisi07}), which produces TRGB luminosities fainter by $\sim0.08$~mag with respect to the standard BaSTI models.

\begin{figure}
 \centering
 \includegraphics[scale=0.5]{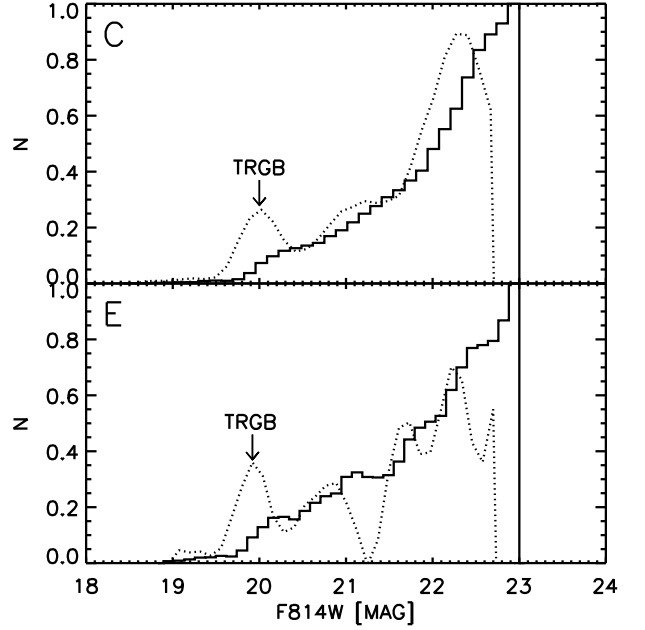}
 \caption{The LF of the RGB of field C (top) and of field E (below) of NGC 6822. The dotted line shows the convolution of the LF with the Sobel kernel, while the arrow marks the location of the TRGB.}
\label{tip_hist}
\end{figure}

To determine the apparent magnitude of the RGB tip in NGC 6822 we adopted the technique suggested by Lee et
al. (1993), which consists of using an edge-detection algorithm -- the Sobel kernel -- with the appropriate
mask [$-$1, $-$2, 0, +2, +1]. Applying this convolution filter to old stellar systems whose AGB population is
sparse, the signature of the tip appears as a sharp peak along the luminosity function (LF) of the RGB.
AGB stars, which are brighter than those on the RGB, do not affect this technique since they have a shorter
evolutionary time scale. To determine the LF of the RGB we considered all stars
belonging to the RGB with $19.0<m<22.5$ and within $3\sigma_m$ from the ridge-line, where $\sigma_m$ is the uncertainty in colour associated to the ridge-line itself at a given magnitude. The LF for the inner field C and, for comparison, for the outer fields E, are shown in Fig.~\ref{tip_hist} along with the result of the convolution of the RGB LF with the Sobel kernel.

Scrutiny of Fig.~\ref{tip_hist} provides the apparent magnitude of the tip in the central and external field as $m^{TRGB}_C= 19.99\pm 0.02$, and $m^{TRGB}_E= 19.90\pm 0.02$ respectively, where the error has been adopted as half of the bin size. As expected, the TRGB for the innermost field is significantly fainter than that of the outermost field. The corresponding colour of the TRGB for field C is readily obtained from the RGB ridge-line, which gives ${\rm (F475W-F814W)_{C}^{TRGB}=3.00\pm 0.02}$.

The estimate of the dereddened TRGB colour was already discussed at the beginning of this section. The
reddening of IC 1613 was adopted from Bernard et al. (\cite{bernard}) as ${\rm E(B-V)=0.026}$ and was
transformed into the appropriate HST bands by using Table~3 of Bedin et al. (\cite{bedin}), who provided the
relationships between the colour excess ${\rm E(B-V)}$ and the extinction coefficients in different HST bands
for cool stars. By interpolating on data in this table we derived ${\rm E(B-V)}=0.56 {\rm E(F475W-F814W)}$ and
${\rm  A_{F814W}= 0.98 E(F475W-F814W)}$. By applying these transformations we obtained the extinction and the absorption of IC 1613 in the HST bands as ${\rm E(F475W-F814W)=0.046}$ and ${\rm A_{F814W}= 0.045}$.

Using the relative reddening found for NGC 6822, we derived the dereddened colour and apparent magnitude of the TRGB for field C:
\begin{equation}
     {\rm  (F475W-F814W)_{0,C}^{TRGB} = 2.36 \pm 0.04}
\label{eq:col_trgb}
\end{equation}
\begin{equation}
     m_{0,C}^{TRGB}= 19.41 \pm 0.03.
\label{eq:mag_trgb}
\end{equation}

In principle, to use the TRGB as distance indicator, an estimate of the average metallicity of the
population of NGC 6822 is required. The upper panel of Fig.~\ref{trgb_models}, based on BaSTI models, shows
the dependence of the luminosity of the TRGB on [Fe/H]. Scrutiny of this figure discloses that in the
metallicity range about $-2.0 \le  {\rm [Fe/H]}\le  -0.8$ and in the age range between about 6 and 12~Gyr, the
absolute magnitude of the tip is nearly constant, ranging from ${\rm M_{F814W}^{TRGB}}=-4.13$ to ${\rm
M_{F814W}^{TRGB}}=-4.05$. We therefore adopted ${\rm M_{F814W}^{TRGB}}=-4.10\pm0.04$. On the basis of this
calibration we estimated the true distance modulus of NGC 6822 as ${\rm (m-M)_0}= m_{0,C}^{TRGB} -{\rm
M_{F814W}^{TRGB} = 23.51\pm 0.05}$. This value is marginally consistent with the distance modulus found at the
beginning of this section. Even if this is not particularly surprising because our first estimate was based on
the variables of IC 1613 as distance indicators, while this second estimate is based on a theoretical
calibration of the TRGB, nonetheless, this discrepancy could be regarded as an indication of the metallicity of
the bulk of the population of NGC6822. Indeed, assuming that the distance modulus is ${\rm (m-M)_0= 23.40\pm 0.05}$, it would follow that  ${\rm M_{F814W}^{TRGB}}=-4.07$, and consequently, from Fig.~\ref{trgb_models}, the average metallicity of NGC 6822 would be about ${\rm  [Fe/H]}\sim -1.0$.

To overcome the uncertainty in metallicity, which reflects in the adopted ${\rm M_{F814W}^{TRGB}}$ value and, then, on the distance, we used as a final approach the theoretical TRGB magnitude-colour relationship. The calibration of this parameter is shown in the lower panel of Fig.~\ref{trgb_models}. 
We emphasize that as shown in the figure, in this case the dependence of the TRGB on the age of the stellar population age is almost vanishing while the effect of metallicity is obviously included in the colour. An overall best fit to these data gave
\begin{equation}
     {\rm M_{F814W}^{TRGB}= 0.08\cdot{col}^2 - 0.40\cdot{col} - 3.63 },
\label{eq:cal_trgb}
\end{equation}

\noindent
where ${\rm col=(F475W-F814W)_0}$ and with a $\chi^2$ equal to 0.003.

Using the colour found in equation (\ref{eq:col_trgb}) for field C of NGC 6822, this relationship gives ${\rm M_{F814W,C}^{TRGB}=-4.13 \pm 0.01}$, where the error accounts only for the uncertainties on the TRGB brightness and colour.

Using the TRGB dereddened apparent magnitude of (\ref{eq:mag_trgb}), we obtained a true distance modulus, ${\rm(m-M)_0}=23.54\pm0.05$. The two distance determinations based on two different parameters, i.e. on the luminosity and on the colour of the TRGB,  agree excellently and we conclude that the distance modulus of NGC 6822 is ${\rm(m-M)_0}=23.53\pm0.05$.

\section{Conclusions}

We carried out an analysis of the dIrr NGC 6822 using photometric data collected with the ACS detector onboard HST. The data analysis was performed using the photometry package software DAOPHOT, supported by ALLFRAME. We obtained very accurate CMDs for three distinct fields of view. In our analysis the CMDs corresponding to the two outermost fields were considered as a unique CMD, because they overlap perfectly. The results obtained in this paper are summarized as follows:

\begin{itemize}

\item We studied the reddening and first investigated the possible differential reddening, which we estimated to be $ {\rm \Delta E(F475W-F814W)= 0.12\pm 0.02}$. Then we evaluated the colour excess ${\rm E(B-V)}$ by a comparison with the CMD of another dIrr, IC 1613, and found $ {\rm E(B-V)_C=0.37\pm 0.02}$ and $ {\rm E(B-V)_E=0.30\pm 0.032}$ for the innermost and outermost fields;
     
\item We estimated the distance modulus with a new and never before used method, the tip of the RGB. This yielded a theoretical calibration of this standard candle based on the TRGB brightness - colour relationship, which we applied to our data set. The result is $ {\rm (m-M)_0=23.54\pm 0.05}$. Although higher by about 0.14~mag, this distance estimate agrees -- within the quoted uncertainty -- with the distance measurement
provided by Feast et al.~(2012). However, it is also worth noting that the measurement by Feast et al. (2012) was based on the assumption that the LMC distance modulus is equal to 18.50. Accordingly, when accounting in the error budget for the still substantial uncertainty (at least of the order of 0.05 mag)  in the adopted LMC distance (Tammann et al.~2006), it is clear that the distance estimates based on classical Cepheids and the TRGB method are quite consistent.

\end{itemize}

\begin{acknowledgements}

We warmly thank our referee for her/his pertinent and very helpful suggestions. SC is grateful for financial support from PRIN-INAF 2009 "Formation and Early Evolution of Massive Star Clusters" (PI: R. Gratton)
and PRIN-INAF 2011 "Multiple Populations in Globular Clusters: their role in the Galaxy assembly" (PI: E. Carretta). Support for this work was provided by the IAC (grants, 310394, 301204), the Education and Science Ministry of Spain (grants AYA2010-16717).
      
\end{acknowledgements}


\begin{thebibliography}{}
   \bibitem[1884]{barnard} Barnard, E. E. 1884, Sideral Messenger, 3, 254
   
   \bibitem [2005]{bedin} Bedin, L. R., Cassisi, S., Castelli, F., Piotto, G., Anderson, J., Salaris, M., Momany, Y., \& Pietrinferni, A. 2005, MNRAS, 357, 1038
   
\bibitem[2008]{bellazzini} Bellazzini, M.  2008, Mem. Soc. Astr. It., 79, 440

   \bibitem [2010]{bernard} Bernard, E. J., Monelli, M., Gallart, C., Aparicio, A., Cassisi, S., Drozdovsky, I., Hidalgo, S. L., Skillman, E. D., \& Stetson, P. B. 2010, ApJ, 712, 1259
   
   \bibitem[2011]{Cannon} Cannon, J. M., Walter, F., Armus, L., Bendo, G. J., Calzetti, D., Draine, B. T., Engelbracht, C. W., Helou, G., Kennicutt, R. C., Jr., Leitherer, C., Roussel, H., Bot, C., Buckalew, B. A., Dale, D. A., de Blok, W. J. G., Gordon, K. D., Hollenbach, D. J., Jarrett, T. H., Meyer, M. J., Murphy, E. J., Sheth, K., \& Thornley, M. D. 2011, ApJ, 652, 1170
   
   \bibitem [2007]{cassisi07} Cassisi, S., Potekhin, A. Y., Pietrinferni, A., Catelan, M., \& Salaris, M. 2007, ApJ, 661, 1094
 
 \bibitem[2012]{cs:2012} Cassisi, S., \& Salaris, M. 2012, "Old stellar populations: how to study the fossil record of galaxy formation", Wiley-VCH eds., {\sl in press}
 
   \bibitem [2005]{cioni} Cioni, M.-R. L., \& Habing, H. J. 2005, A\& A, 429, 837
   
   \bibitem [1990]{dacosta} Da Costa, G. S., \& Armandroff, T. E. 1990, AJ, 100, 162
   
   \bibitem[2000]{deblok} de Blok, W. J. G., \& Walter, F. 2000, AJ, 537, L95
   
   \bibitem [2012]{feast} Feast, M. W., Whitelock, P. A., Menzies, J. W., \& Matsunaga, N. 2012, MNRAS, 421, 2998
   
   \bibitem[1996]{gallart} Gallart, C., Aparicio A., \& Vilchez, J. M. 1996, AJ, 112, 1928
   
   \bibitem[2006]{gieren} Gieren, W., Pietrzynski, G., Nalewajko, K., Soszynski, I., Bresolin, F., Kudritzki, R.-P., Minniti, D., \& Romanowsky, A. 2006, ApJ, 647, 1056
   
   \bibitem[2012]{kacharov} Kacharov, N., Rejkuba, M., \& Cioni, M.-R. L. 2012, A\& A, 537, 108
   
   \bibitem [1993]{lee} Lee, M. G., Freedman, W., \& Madore, B. F. 1993, ApJ, 417, 553
   
   \bibitem[1998]{mateo} Mateo, M. 1998, ARA\& A, 36, 435
   
   \bibitem [1995]{massey} Massey, P., Armandroff, T. E., Pyke, R., Patel, K., \& Wilson, C. D. 1995, AJ, 110, 2715
   
   \bibitem [2004]{pietrinferni04} Pietrinferni, A., Cassisi, S., Salaris, M., \& Castelli, F. 2004, ApJ, 612, 168
   
   \bibitem [2006]{pietrinferni06} Pietrinferni, A., Cassisi, S., Salaris, M., \& Castelli, F. 2006, ApJ, 642, 797
   
   \bibitem [2007]{rizzi} Rizzi, L., Tully, R.~B., Makarov, D., Makarova, L., Dolphin, A.~E., Sakai, S., Shaya, E.~J. 2007, ApJ, 661, 815

\bibitem[1998]{sc:98} Salaris, M. \& Cassisi, S. 1998, MNRAS, 298, 166

\bibitem[2005]{sc:05} Salaris, M. \& Cassisi, S. 2005, "Evolution of stars and stellar populations", Wiley-VCH eds, pp. 400

   \bibitem[2011]{schlafly} Schlafly, E. F., \& Finkbeiner, D. P. 2011, ApJ, 737, 103
   
   \bibitem[1998]{schlegel} Schlegel, D. J., Finkbeiner, D. P., \& Davis, M. 1998, ApJ, 500, 525
   
   \bibitem[2012]{sibbons} Sibbons, L.F., Ryan, S. G., Cioni, M.-R. L., Irwin, M., \& Napiwotzki, R. 2012, A\& A, 540, 135
   
   \bibitem [2005]{sirianni} Sirianni, M., Jee, M. J., Benatez, N., Blakeslee, J. P., Martel, A. R., Meurer, G., Clampin, M., et al.  2005, PASP, 117, 1049
   
   \bibitem[1989]{skillman} Skillman, E. D., Terlevich, R., \& Melnick, J. 1989, MNRAS, 240, 563
   
   \bibitem[1987]{daophot} Stetson, P. B. 1987, PASP, 99, 191
   
   \bibitem [1994]{allframe} Stetson, P. B. 1994, PASP, 106, 250
   
   \bibitem [2012]{tammann} Tammann, G. A., \& Reindl, B. 2012, Astrophysics and Space Science, 341, 3
   
   \bibitem[2008]{tammann08} Tammann, G. A., Sandage, A.,  \& Reindl, B. 2008, ApJ, 679, 52
   
   \bibitem[2001]{tolstoy} Tolstoy, E., Irwin, M. J., Cole, A. A., Pasquini, L., Gilmozzi, R., \& Gallagher, J. S. 2001, MNRAS, 327, 918
   
   \bibitem[1998]{vandenberg} van den Bergh, S. 1998, in Galaxy Morphology and Classification (Cambridge: Cambridge University Press)
   
   \bibitem[2001]{venn} Venn, K. A., Lennon, D. J., Kaufer, A., et al. 2001, ApJ, 547, 765

\bibitem[2012]{whitelock} Whitelock, P.A., Menzies, J.W., Feast, M.W., Nsengiyumva, F., \& Matsunaga, N. 2012, MNRAS, {\sl in press}, arXiv:1210.3695
   
  
\end{thebibliography}
\end{document}